\documentclass[aps,prl,twocolumn,showpacs,superscriptaddress,groupedaddress]{revtex4-1}

\usepackage{amsmath}
\usepackage{amssymb}
\usepackage{graphicx}
\usepackage{bm}
\usepackage{times}

\begin{document}

\title{Simultaneous Measurement of Complementary Observables with
  Compressive Sensing}

\author{Gregory A. Howland}
\email{ghowland@pas.rochester.edu}
\author{James Schneeloch}
\author{Daniel J. Lum}
\author{John C. Howell}
\affiliation{Department of Physics and Astronomy, University of Rochester \\
500 Wilson Blvd, Rochester, NY 14618}

\begin{abstract}
  The more information a measurement provides about a quantum system's
  position statistics, the less information a subsequent measurement
  can provide about the system's momentum statistics. This information
  trade-off is embodied in the entropic formulation of the uncertainty
  principle. Traditionally, uncertainty relations correspond to
  resolution limits; increasing a detector's position sensitivity
  decreases its momentum sensitivity and vice-versa. However, this is
  not required in general; for example, position information can
  instead be extracted at the cost of noise in momentum. Using random,
  partial projections in position followed by strong measurements in
  momentum, we efficiently determine the transverse-position and
  transverse-momentum distributions of an unknown optical field with a
  single set of measurements. The momentum distribution is directly
  imaged, while the position distribution is recovered using
  compressive sensing. At no point do we violate uncertainty
  relations; rather, we economize the use of information we obtain.
\end{abstract}


\maketitle 


Measurements on quantum systems are always constrained by uncertainty
relations. Localizing a particle in one observable, such as position,
imparts a disturbance that makes a following measurement of a
complementary observable, such as momentum, unpredictable. Such
\emph{strong}, projective measurements are often said to ``collapse''
the quantum wavefunction. For example, in Young's double slit
experiment, it is not possible to detect through which slit particles
pass (position) while also observing interference fringes in the far
field (momentum) \cite{bohr:1928}.

Consequently, the statistics of complementary observables are usually
measured separately; an ensemble of identically prepared particles is
directed to a position detector and a different, similarly prepared
ensemble is directed to a momentum detector. If a detector instead
measures both observables simultaneously with strong measurements, its
position resolution $\Delta_x$ and momentum resolution $\Delta_k$ are
bounded by Heisenberg's uncertainty relation, $\Delta_x \Delta_k \ge
1/2$. In its most basic form, a Shack-Hartmann wavefront sensor is an
example of this kind of detector \cite{platt:2001}.

Though this resolution limitation applies to strong measurements, it
is not true in general. Here, the uncertainty principle implies an
information exclusion principle \cite{maassen:1988,hall:1995}; the
more information a detector gives about position, the less information
it can provide about momentum and vice-versa. With a single, carefully
designed experiment, one can simultaneously recover the statistics of
both observables at arbitrary resolution. This has been demonstrated,
albeit very inefficiently, with weak measurement
\cite{lundeen:2011,kocsis:2011,dressel:2013}.

\begin{figure}[h]
  \includegraphics[scale=0.4]{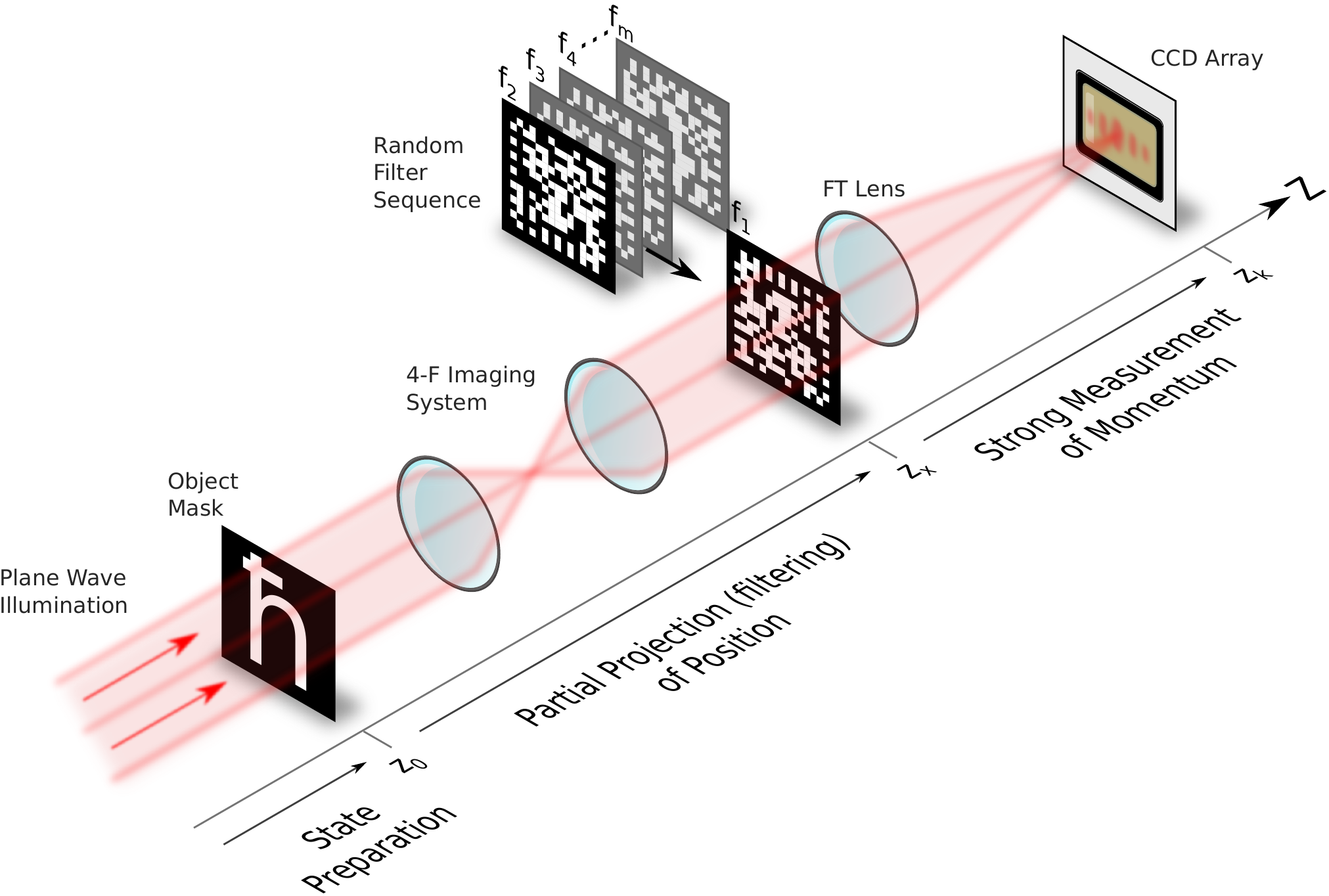}
  \caption{\textbf{Experimental setup for simultaneous position and
      momentum imaging.} A state is prepared by illuminating an object
    mask at $z=z_0$ with a plane wave from an attenuated, HeNe
    laser. The field is imaged at $z=z_x$ where it is sequentially
    filtered by a series of $M$, $256\times 256$ pixel, random, binary
    filters $\bm{F_i}$. Each filter partially projects the state by
    blocking about half of the position elements. A cooled CCD array
    in the focal plane $z=z_k$ of a Fourier transforming lens records
    the momentum distribution for each filtered state. Position
    information is mapped to the total optical power passing each
    filter; this measures the correlation between the position
    intensity distribution at $z_x$ with the current filter. Because
    the filters do not strongly localize the photons' position, the
    momentum distribution is directly recovered by averaging the CCD
    images. The position distribution is reconstructed using
    compressive sensing techniques such that an $N$-pixel position
    image requires $M<<N$ filters.}
  \label{fig:setup}
\end{figure}

In this Letter, we efficiently obtain the transverse-position and
transverse-momentum distributions of optical photons from a single set
of measurements at high resolution. We sequentially perform a series
of random, partial projections in position followed by strong
projective measurements of the momentum. The partial projections
efficiently extract information about the photons' position
distribution at the cost of injecting a small amount of noise into
their momentum distribution. This allows the momentum distribution to
be directly observed on a charge-coupled device (CCD) camera. The
position distribution is recovered using a computational technique
called compressive sensing (CS) \cite{donoho:2006}.

Consider an optical field at plane $z=z_0$ with transverse, complex
amplitude $\psi(\vec{x})$, where $z$ is the propogation direction and
$\vec{x}=(x,y)$ are transverse, spatial coordinates. The field also
has momentum amplitude $\psi(\vec{k})$ which is related to
$\psi(\vec{x})$ by a Fourier transform, with $\vec{k}=(k_x,k_y)$.

To measure the position field intensity $|\psi(\vec{x})|^2$, one could
raster scan a small pinhole through the transverse plane at
$z=z_0$. The fractional power passing through the pinhole as a
function of its position reveals the image. From a quantum
perspective, this process constitutes a \emph{strong} projective
position measurement; the pinhole localizes the position of photons
passing through it and their subsequent momenta are random. From a
classical optics perspective, the pinhole acts like a spatial filter;
light passing through the pinhole diffracts evenly in all
directions. In either case, information about the original field's
momentum $\psi(\vec{k})$ is lost. Note that one could instead choose
to measure the momentum distribution $|\psi(\vec{k})|^2$ by performing
a similar scan in the focal plane of a Fourier transforming
lens. Here, position information would instead be forfeit.

In our approach (Fig. \ref{fig:setup}), we perform a series of
partially projective measurements of position followed by strong
measurements of momentum. We first prepare a transverse, photonic
state $\psi(\vec{x})$ by illuminating an object mask with a collimated
laser. We image this field at plane $z = z_x$ with a 4F imaging
system. Here we sequentially perform partial projections of
$\psi(\vec{x})$ by filtering it with a series of $M$ binary amplitude
masks $f_i(\vec{x})$. Each mask consists of a random, $N$-pixel
pattern, where each pixel either fully transmits or fully obstructs
with equal probability. Note that the total optical power passing the
$i^{\text{th}}$ filter gives the correlation between that filter and
the position intensity distribution $|\psi(\vec{x})|^2$. In this way,
a small amount of information about the position distribution is
extracted without localizing the field.  The filtered state
$\tilde{\psi}_i(\vec{x}) = \psi(\vec{x})f_i(\vec{x})$ then passes
through a Fourier transforming lens to a CCD array in the lens' focal
plane at $z = z_k$. The CCD records $M$ images of the momentum
distribution of the filtered field $|\tilde{\psi}_i(\vec{k})|^2$, one
for each filter. This set of images contains information about both
$\psi$'s position and momentum.

The momentum distribution $|\tilde{\psi}(\vec{k})|^2$ is recovered
directly from the CCD images by simple averaging such that
\begin{equation}
 |\psi(\vec{k})|^2 = \langle |\tilde{\psi}_i(\vec{k})|^2\rangle,
 \label{eq:avg}
\end{equation}
where angled brackets indicate an average over all filters. This is
made possible by the surprising fact that
$|\tilde{\psi}_i(\vec{k})|^2$ is a good approximation to
$|\psi(\vec{k})|^2$, even though $|\tilde{\psi}_i(\vec{x})|^2$ is
missing half of its coefficients.

\begin{figure}[t]
  \includegraphics[scale=0.4]{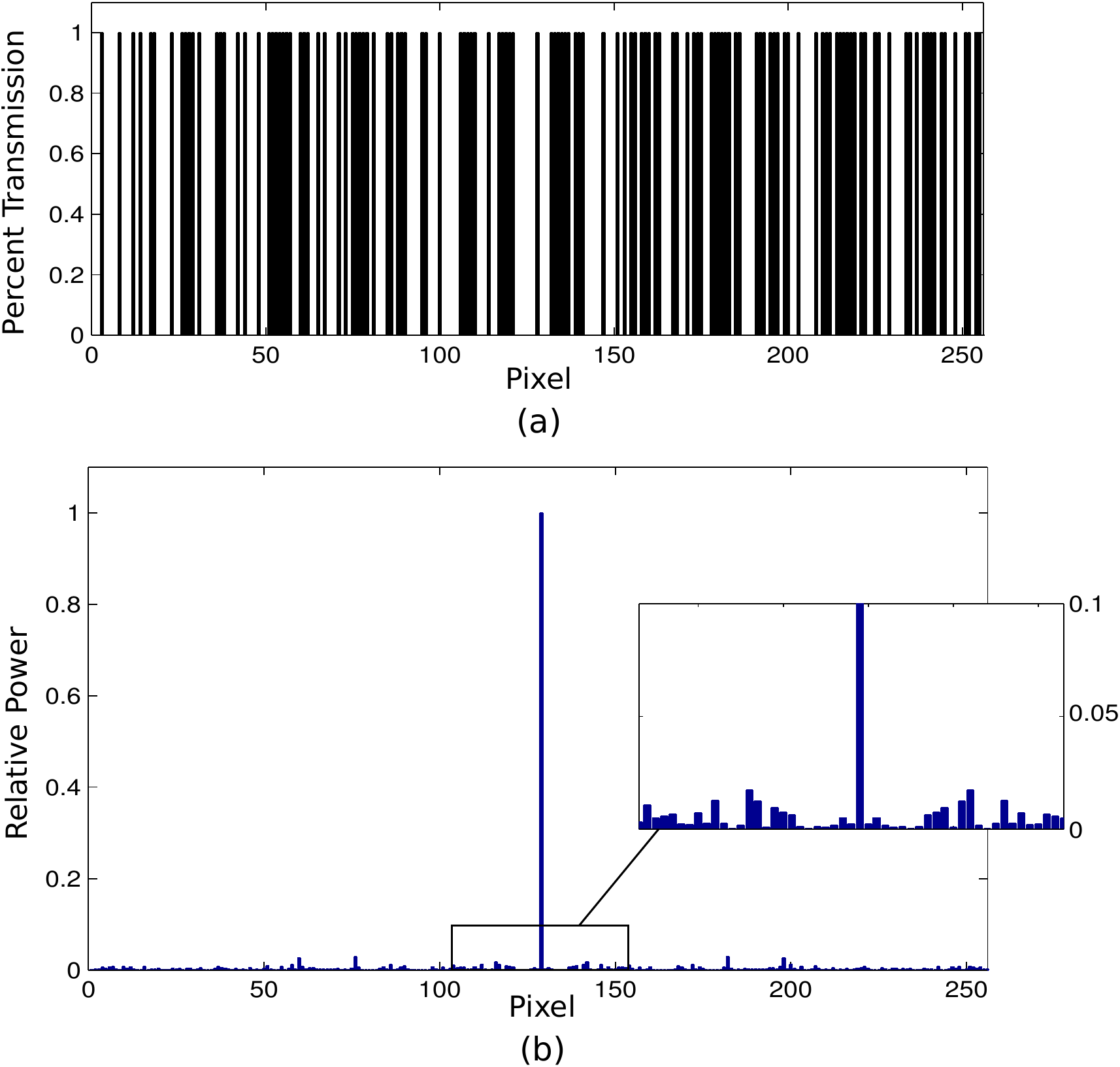}
  \caption{{\bf Discrete Fourier transform of a 256 pixel, 1D random,
      binary pattern.} (a) gives a random, binary 1D filter function
    where a value 1 is fully transmitting and a value 0 is fully
    obstructing. (b) shows the relative power spectrum of its Fourier
    transform, where the zero momentum term is scaled to unity. The
    noise floor is a factor $\sqrt{2/N}$ weaker in amplitude in both
    its mean and standard deviation. The same relationship holds for a
    2D filter.}
  \label{fig:pat}
\end{figure}

By the convolution theorem of Fourier optics \cite{goodman:2005}, the
filtered $i^{\text{th}}$ momentum distribution is found by convolving
the Fourier transforms of $\psi(\vec{x})$ and $f_i(\vec{x})$ such that
\begin{equation}
  |\tilde{\psi}_i(\vec{k})|^2 = |\psi(\vec{k}) \ast f_i(\vec{k})|^2,
  \label{eq:conv}
\end{equation}
where $\ast$ denotes convolution. To understand the filter's effect on
$\psi(\vec{k})$, we must consider its Fourier transform
(Fig. \ref{fig:pat}).

At high resolution, each transmitting filter pixel is approximately a
displaced Dirac delta function (Fig. \ref{fig:pat}a) with unit
amplitude. The Fourier transform of each delta function is a plane
wave propagating at an angle proportional to its displacement from the
origin. At $\vec{k} = (0,0)$, these plane waves add in phase,
producing a sharp peak. For $\vec{k} \ne (0,0)$, each plane wave is
equally likely to provide a negative or positive contribution. The
coefficients therefore follow a random, complex Gaussian distribution
\cite{mccrea:1940}. A filter's Fourier transform is approximately a
Dirac delta function at zero momentum riding a small noise floor a
factor $\sqrt{2/N}$ weaker (Fig. \ref{fig:pat}b)
\begin{equation}
  f_i(\vec{k}) \propto \delta(\vec{k})+\sqrt{\frac{2}{N}}\phi_i(\vec{k}).
\label{eq:filt}
\end{equation} 
Values for $\phi_i(\vec{k})$ follow a random, complex Gaussian white
noise distribution, with real and imaginary parts of zero mean and
standard deviation $1/\sqrt{2}$.

\begin{figure}[t]
\includegraphics[scale=0.35]{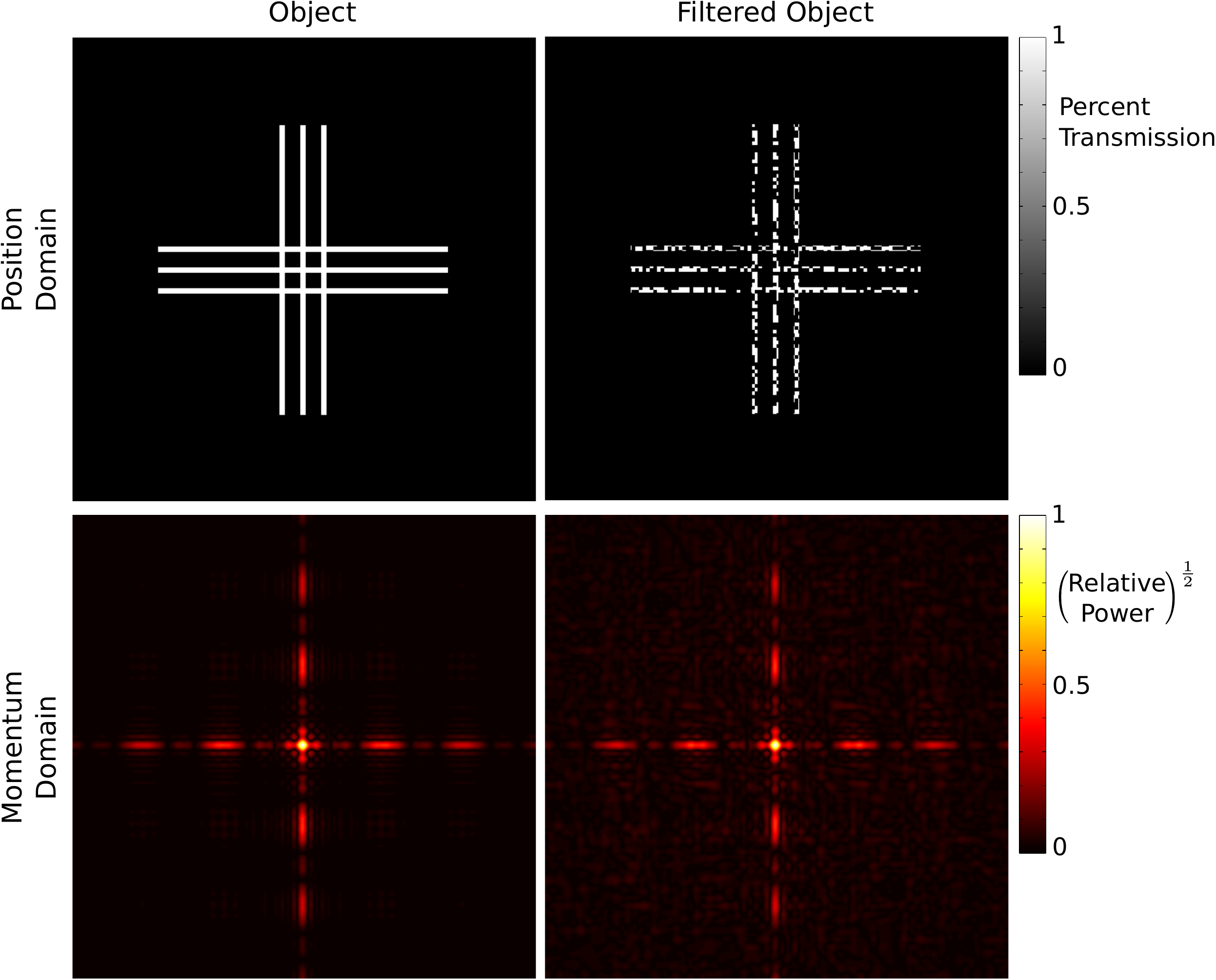}
\caption{\textbf{Partial projections of $\bm{\psi}$.} The figure
  simulates the effect of filtering a triple slit object with a
  $128\times 128$ pixel random, binary pattern. The momentum images
  are given as relative powers; where their maximum value is scaled to
  unity. A square-root color mapping emphasizes weaker momentum
  values. The filtered momentum distribution is a slightly noisy
  version of the true momentum distribution.}
\label{fig:pert}
\end{figure}

Because a convolution with a delta function simply returns the
original function, we expect $\tilde{\psi}_i(\vec{k}) \approx
\psi(\vec{k})$ with a small amount of noise
(Fig. \ref{fig:pert}). From Eq. \ref{eq:conv} and Eq. \ref{eq:filt},
we find
\begin{align}
  |\tilde{\psi}_i(\vec{k})|^2 &= \mathcal{N} \bigg \{|\psi(\vec{k})|^2 \\
  & +\frac{2\sqrt{2}}{\sqrt{N}}\text{Re}[\psi^\ast(\vec{k})(\psi(\vec{k})\ast\phi_i(\vec{k}))] \nonumber\\
  & +\frac{2}{N}|\psi(\vec{k})\ast\phi_i(\vec{k})|^2\bigg \},
  \nonumber
\end{align}
where $\mathcal{N}$ is a normalizing constant.  The first term is the
desired outcome; the following two terms add noise. For large $N$,
these terms vanish. In the worst case, the signal-to-noise ratio
scales as $\sqrt{N}$. At typical imaging resolutions, such as
$N=256\times 256$ pixels used in this letter, these terms are
weak. When averaged over many patterns, the second term vanishes and
the third term approaches a very small constant
value. Eq. \ref{eq:avg} is therefore recovered up to a constant
offset. The noiseless case is asymptotically approached for increasing
$M$ and $N$. This analysis is closely related to similar problems in
wireless communication \cite{tse:2005}.

Because the filtered momentum distribution is only lightly perturbed,
very little information about the position distribution can be
extracted from each CCD image. To maximize the usefulness of this
information, we turn to compressive sensing \cite{baraniuk:2007,
  baraniuk:2008,romberg:2008}. Compressive sensing \cite{donoho:2006}
is an extremely efficient measurement technique for recovering an
$N$-dimensional signal from $M<<N$ measurements, provided the signal
can be compressed in a known way. The use of outside information, the
prior knowledge that a signal is compressible, is a powerful tool for
economizing measurement. In the past decade, compressive sensing has
taken the signal processing world by storm with applications ranging
from magnetic resonance imaging \cite{lustig:2007} to radio astronomy
\cite{bobin:2008}. More recently, CS has made inroads into the quantum
domain with compressive tomography \cite{gross:2010,cramer:2010,
  shabani:2011}, and entanglement characterization
\cite{howland:2013}. When used for imaging, compressive sensing is
closely related to computational imaging \cite{sun:2013, katz:2009}.

Together, the filters and CCD implement a single-pixel camera for the
position distribution. The single-pixel camera is the textbook example
of compressive sensing and has been extensively investigated
\cite{baraniuk:2008,takhar:2006}. Consider the total power $Y_i$
striking the CCD while filtering with $f_i$, obtained by integrating
the $i^{\text{th}}$ momentum image $|\tilde{\psi}_i(\vec{k})|^2$ over
all CCD pixels. The CCD now acts as a single-element power meter. The
value $Y_i$ is a correlation between the position intensity
$|\psi(\vec{x})|^2$ and the $i^{\text{th}}$ filter.

These correlations are concisely represented by the series of linear
equations
\begin{equation}
\bm{Y} = \bm{FX}.
\label{eq:meas}
\end{equation}
Here, $\bm{F}$ is an $M\times N$ sensing matrix whose $i^{\text{th}}$
row is a 1D reshaping of the $i^{\text{th}}$ filter function. $\bm{X}$
is an $N$-dimensional vector representing a 1D reshaping of the
unknown position distribution $|\psi(\vec{x})|^2$, discretized to the
same resolution as the filters.

The correlations can be used to iteratively recover $\bm{X}$ by taking
a weighted sum of the filter functions
\begin{equation}
  \bm{X} = \frac{1}{M}\sum_{i=1}^{M}\bm{Y}_i\bm{F}_i,
\end{equation}
but many measurements are required ($M\ge N$)
\cite{welsh:2013}. Instead, given some reasonable assumptions,
compressive sensing dramatically reduces the requisite number of
measurements ($M<<N$).

When $M<<N $, Eq. \ref{eq:meas} is under-determined; there are many
possible $\bm{X}$ consistent with $\bm{Y}$. CS posits that the correct
$\bm{X}$ is the one that is sparsest (has the fewest number of
non-zero elements) in a representation where $\bm{X}$ is
compressible. This $\bm{X}$ is found by solving the regularized
least-squares optimization problem
\begin{equation}
\min_{\bm{X}} \frac{\mu}{2}||\bm{Y}-\bm{FX}||_2^2 + TV(\bm{X}),
\label{eq:obj}
\end{equation}
where for example $||\bm{q}||_2^2$ is the $\ell_2$ norm (Euclidean
norm) of $\bm{q}$ and $\mu$ is a constant. The first penalty is a
least-squares term that is small when $\bm{X}$ is consistent with the
correlation vector $\bm{Y}$. The second penalty $TV(\bm{X})$ is the
signal's total variation,
\begin{equation}
  TV(\bm{X}) = \sum_{\text{adj. }i,j}|\bm{X}_i-\bm{X}_j|,
\end{equation}
where indices ${i,j}$ run over all pairs of adjacent pixels in
$\bm{X}$ so that $TV(\bm{X})$ is just the $\ell_1$ norm of $\bm{X}$'s
discrete gradient. 

If a signal's total variation is large, values of adjacent pixels vary
wildly, indicating a noisy, unstructured signal. Conversely, when a
signal's total variation is small, values for adjacent pixels are
strongly correlated, indicating structure consistent with a real
image. Put more plainly, we seek the signal with the fewest edges
consistent with our measurements; this leverages compressibility in
$\bm{X}$'s gradient. Total variation minimization has proven extremely
effective for compressive imaging; exact recovery of $\bm{X}$ is
possible with $M$ as low as a few percent of $N$
\cite{candes:2006}. In addition to sub-Nyquist sampling, CS has been
shown to give a higher signal-to-noise ratio than raster- or
basis-scan \cite{candes:2008}.

\begin{figure}[t]
\includegraphics[scale=0.5]{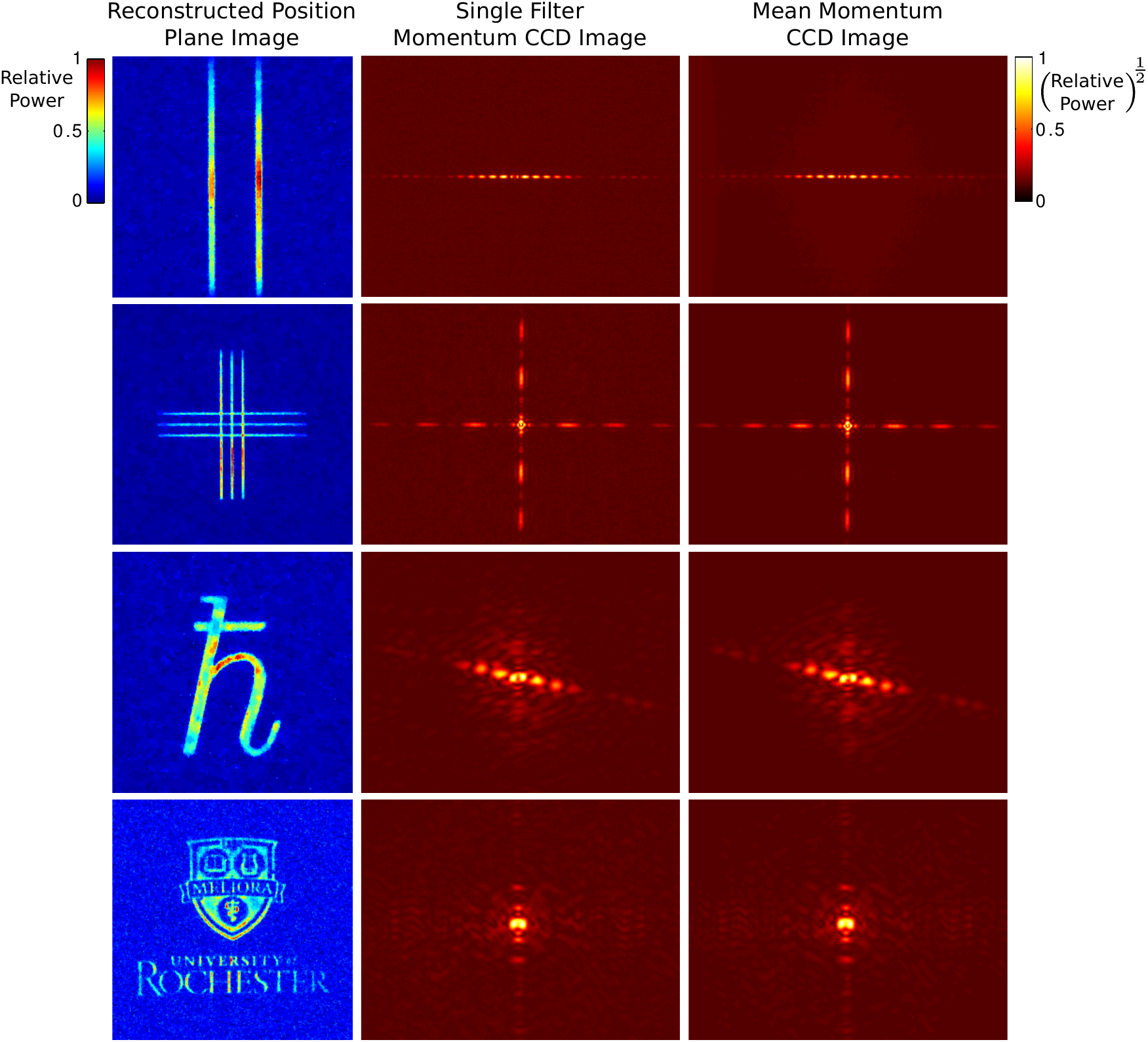}
\caption{\textbf{Recovered position and momentum images for four
    objects.} A double slit, triple slit, and $\hbar$ were
  reconstructed at $N=256\times 256$ resolution from only $M=6553$
  filters; the university logo used $M=32768$ filters.  No additional
  post-processing has been performed; position images are those
  returned by the reconstruction algorithm and momentum images are the
  recorded single or mean CCD images.}
\label{fig:results}
\end{figure}

We tested our technique on four objects: a double slit, a triple slit,
the character $\hbar$, and the University of Rochester logo
(Fig. \ref{fig:results}). The object and filter masks were introduced
using computer controlled spatial light modulators, which can change
patterns at typical video speeds up to $60$ Hz. The filter spatial
resolution was $N=256\times 256$ pixels. The random filter functions
were rows of a randomly permuted, zero-shifted Hadamard matrix
\cite{li:2011}. This allows $\bm{Y}=\bm{FX}$ to be efficiently
computed by a fast transform when solving Eq. \ref{eq:obj}.

The CCD was a cooled, $12$ bit, $1376\times1040$ pixel sensor. The
exposure time for each CCD image was $10$ ms. The average optical
power incident on the CCD was of order $10$ pW. For $10$ ms exposures,
each CCD pixel had dark noise $50 \pm 10$ in arbitrary power units of
$0$ to $4096$. When integrating the CCD image to produce the
correlation vector $\bm{Y}$, this value was subtracted. Momentum
images are those recorded directly by the camera; no post-processing
is performed beyond averaging over all images.

\begin{figure}[t]
\includegraphics[scale=0.55]{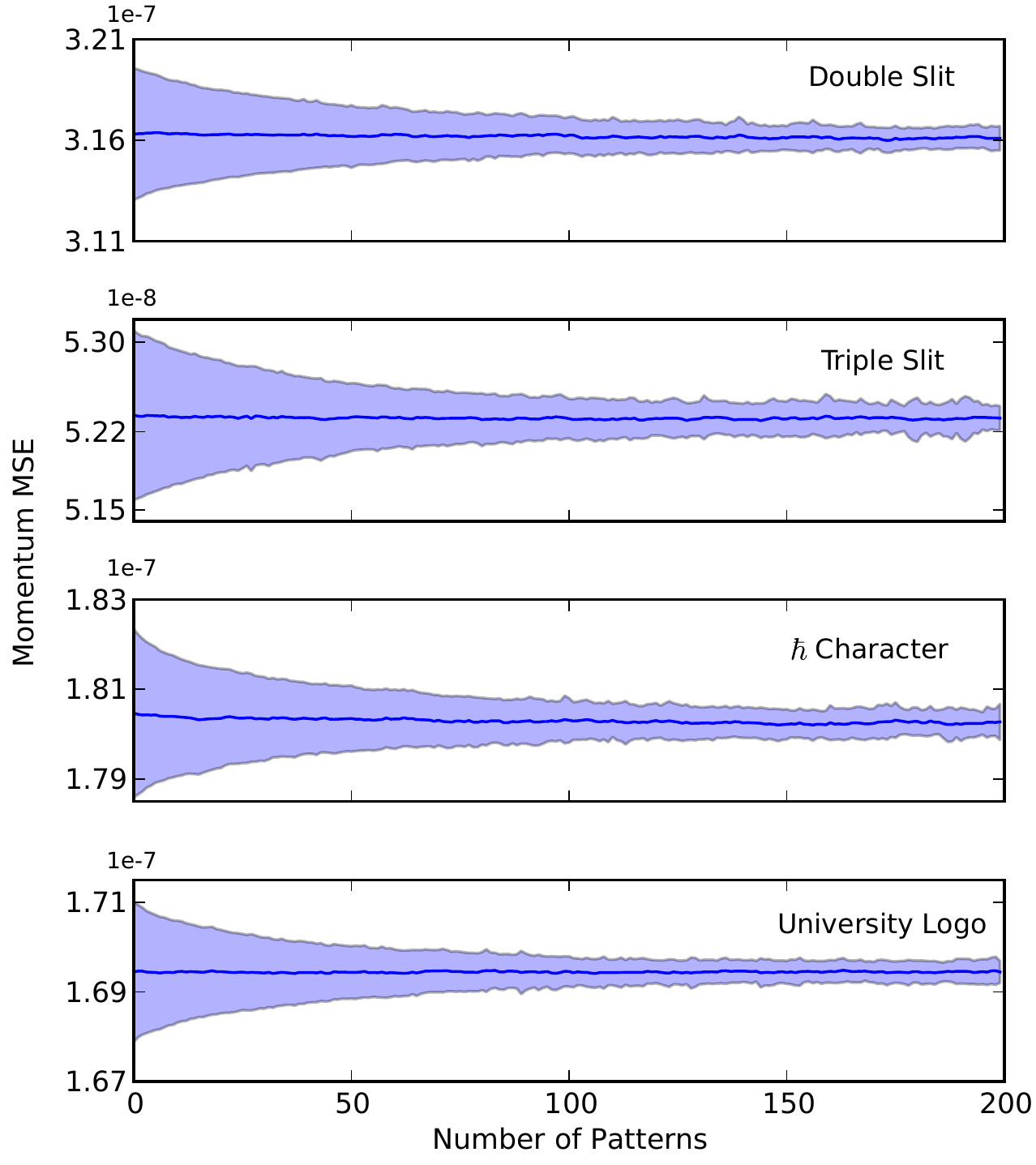}
\caption{{\bf Simulated Momentum MSE:} The simulated momentum mean
  squared errors (MSE) of the four objects used in the experiment are
  given as a function of the number of random patterns $M$. The shaded
  region encloses one standard deviation above and below the average
  MSE for 100 trials. The MSE rapidly approaches a small constant
  value as the second term of Eq. 4 vanishes. Even for a single
  pattern, the MSE is at least of order $10^{-7}$.}
\label{fig:mse}
\end{figure}

For the double slit, triple slit, and character objects, $M=0.1N=6553$
filters were used; for the university logo, $M=0.5N=32768$ filters
were used. These correspond to total exposure times of $65.5$ sec and
$327.7$ sec respectively. Note that Nyquist sampling would require $N$
measurements; for most objects we undersample by an order of
magnitude. The requisite $M$ depends both on object complexity and the
chosen objective function (Eq. \ref{eq:obj}), sensing matrix, and
solving algorithm. We have chosen conservatively large $M$ to produce
high quality images. The dependence of image quality on $M$ is
extensively researched; for example see Refs. \cite{candes:2008,
  romberg:2008, baraniuk:2008}.

The position distributions were reconstructed by solving
Eq. \ref{eq:obj} using the TVAL3 solver \cite{li:2009}. Values of
$\mu$ ranged from $2^{10}$ to $2^{14}$. Such large $\mu$ strongly
favors the least squares penalty of Eq. \ref{eq:obj} such that it is
effectively a constraint.

In all cases, our technique recovered high fidelity position and
momentum distributions. Even momentum images for a single filter are
good approximations to the true distribution; these are further
improved by averaging.

To show the accuracy of our technique, Fig. \ref{fig:mse} gives the
mean squared errors (MSE) of the momentum images for $100$ simulations
of the objects used in the experiment as a function of increasing
$M$. Even for a single pattern, the MSE is at least of order
$10^{-7}$. Averaging over an increasing number of patterns, the middle
term of Eq. 4 vanishes and the MSE approaches a constant. This occurs
within a few hundred patterns, well before the requisite $M$ for
recovering the position image.

We have demonstrated an efficient technique for measuring the
probability distributions of complementary observables from a single
set of measurements. Beyond fundamental interest, we anticipate that
our approach will be useful for a wide variety of quantum and
classical sensing tasks, including continuous quantum measurement
\cite{jacobs:2006}, high-dimensional entanglement characterization
\cite{howland:2013}, wavefront sensing, and phase retrieval
\cite{fienup:1982}. We strongly emphasize that our technique does not
violate the uncertainty principle; at no point does a single detection
event give precise information about both position and
momentum. Instead, each detection event gives some information about
both domains. Our approach economizes the use of this
information. More broadly, our system exemplifies a trend in sensing
away from traditional strong projective measurements and raster scans
which scale poorly to large dimensions. Novel techniques based on
compressive sensing, weak measurement, and other unorthodox strategies
are necessary to overcome these limitations.

This work was supported by AFOSR grant FA9550-13-1-0019 and DARPA DSO
InPho grant W911NF-10-1-0404.


\bibliography{refs}

\section{Supplemental Material}

\section{Theory of Random, Binary Partial Projections}

Here we model partial projective measurements in position as random,
binary, pixellated filter functions $f_{i}(\vec{x})$, where $i$ is an
index for each filter function. We examine the statistics of such
random filter functions in transverse-position and transverse-momentum
space, and discuss their effect on the momentum probability
distribution of an object field, $|\psi(\vec{k})|^{2}$.

\subsection{Fourier Transforms of Random Binary Patterns}

We model the random filter functions in position space as a sum of
Dirac delta functions arranged on a regular lattice, multiplied either
by unity with probability $P$ or zero with probability $1-P$;
\begin{equation}\label{posSpaceFilter}
  f_{i}(\vec{x}) = \sum_{\ell,m} a_{\ell,m}^{(i)} \delta(x-g\ell,y-gm).
\end{equation}
Here we have a square lattice of $N$ points centered at
$\vec{x}=(0,0)$ with spacing $g$ in the $x$ and $y$ directions. The
weights $a_{\ell}^{(i)}$ take values zero or unity according to
$P$. Taking the Fourier transform of $f_{i}(\vec{x})$, we find
\begin{equation}\label{momAmp}
  f_{i}(\vec{k}) = \frac{1}{2\pi}\sum_{\ell,m} a_{\ell,m}^{(i)} e^{-ig(k_x \ell +k_y m)}.
\end{equation}

To model these filter functions in momentum, we make the following
assumptions. First, since $N$ is large, and each weight has
probability $P$ of being unity and is otherwise zero, we represent
$f_{i}(\vec{k})$ as a sum of $NP$ unit phasors. Therefore,
$f_{i}(\vec{k}=0)$ is $NP/2\pi$.

Second, since the weights are randomly distributed, we represent
$f_{i}(\vec{k}\ne 0)$ as a sum of $NP$ random phasors. This sum is
well described by a two-dimensional, complex random walk with unit
step size \cite{mccrea:1940,tse:2005}. Therefore, nonzero frequency
components have zero-mean and average square-magnitude $NP/(2\pi)^2$.
When $NP$ is large, it follows from the central limit theorem that the
non-zero frequency components are described by a circularly symmetric,
complex Gaussian distribution with real and imaginary widths
$\sigma$. The momentum amplitude distribution of a typical filter is a
sharply peaked function centered at the origin.


$f_{i}(\vec{k})$ can now be written as a weighted sum of a Dirac delta
function $\delta(\vec{k})$ and a noise function $\phi_{i}(\vec{k})$
whose phases vary uniformly;
\begin{equation}\label{toy}
  f_{i}(\vec{k})\approx \alpha \delta(\vec{k}) + \beta \phi_{i}(\vec{k}),
\end{equation}
where $\alpha$ and $\beta$ are parameters we estimate from the
pattern-averaged values of $f_{i}(\vec{k})$ and
$|f_{i}(\vec{k})|^{2}$. Values for $\phi_i(\vec{k})$ follow a random,
complex Gaussian white noise distribution, with real and imaginary
parts of zero mean and $\sigma = 1/\sqrt{2}$.

Knowing $f_{i}(\vec{k}=0)=NP/2\pi$ gives $\alpha = NP/2\pi$. Knowing
that $\langle f_{i}(\vec{k}\neq0)\rangle=0$, and that
$\langle|f_{i}(\vec{k}\neq0)|^{2}\rangle= NP/(2\pi)^{2}$, we have that
$\beta = \sqrt{NP}/2\pi$, where $\langle\cdot\rangle$ is an average
over many filter functions. Therefore, a viable model for
$f_{i}(\vec{k})$ is
\begin{equation}\label{toyapprox}
  f_{i}(\vec{k}) \approx \frac{NP}{2\pi}\delta(\vec{k}) + \frac{\sqrt{NP}}{2\pi}\phi_{i}(\vec{k}).
\end{equation}

As mentioned previously, our model assumes that $f_{i}(\vec{k})$ is a
sum of $NP$ random phasors. However, since we are sampling these
phasors without replacement, $\langle|f_{i}(\vec{k}\neq0)|^{2}\rangle$
is actually less than $NP/(2\pi)^{2}$; $NP/(2\pi)^{2}$ is a
conservative estimate which will over-estimate the perturbation to
$|\psi(\vec{k})|^{2}$ due to $f_{i}(\vec{k})$.  


\subsection{Effect of a random pattern on momentum distribution}
Let $\psi(\vec{x})$ be the unperturbed position amplitude of the
field. After interacting with filter $f_{i}(\vec{x})$, the perturbed
position amplitude is $\tilde{\psi}(\vec{x}) = f_{i}(\vec{x})\psi
(\vec{x})$. Therefore, the perturbed momentum amplitude is
$\tilde{\psi}(\vec{k})=\psi(\vec{k})\ast f_{i}(\vec{k})$, where $\ast$
denotes convolution.

Using Eq. \eqref{toyapprox}, we find
\begin{equation}
  \tilde{\psi}(\vec{k}) =\frac{NP}{2\pi}(\psi(\vec{k})\ast \delta(\vec{k})) + \frac{\sqrt{NP}}{2\pi}(\psi(\vec{k})\ast \phi_{i}(\vec{k})).
\end{equation}
Since the first term is a convolution with a delta function, we find
\begin{equation}
  \tilde{\psi}(\vec{k}) =\frac{NP}{2\pi}\psi(\vec{k}) + \frac{\sqrt{NP}}{2\pi}(\psi(\vec{k})\ast \phi_{i}(\vec{k})).
\end{equation}
Taking the modulus square of $\tilde{\psi}(\vec{k})$ gives us the
perturbed momentum distribution
\begin{align}
  |\tilde{\psi}(\vec{k})|^{2}&=\mathcal{N}\bigg[|\psi(\vec{k})|^{2} + \frac{1}{NP}|(\psi(\vec{k})\ast\phi_{i}(\vec{k}))|^{2} + \nonumber\\
  &+
  2\frac{1}{\sqrt{NP}}\mathbf{Re}[\psi^{*}(\vec{k})(\psi(\vec{k})\ast\phi_{i}(\vec{k}))]\bigg],
\end{align}
where $\mathcal{N}$ is a normalization constant.

To see how $|\tilde{\psi}(\vec{k})|^{2}$ compares to the unperturbed
probability distribution $|\psi(\vec{k})|^{2}$, it suffices to know
that $|(\psi(\vec{k})\ast\phi_{i}(\vec{k}))|^{2}$ and
$[\psi^{*}(\vec{k})(\psi(\vec{k})\ast\phi_{i}(\vec{k}))]$ are both of
the order unity. As $NP$ becomes large, $\frac{1}{\sqrt{NP}}$ becomes
small, and $|\tilde{\psi}(\vec{k})|^{2}$ approaches
$|\psi(\vec{k})|^{2}$.

More importantly, we recover $|\psi(\vec{k})|^{2}$ (up to a uniform
constant) from averaging $|\tilde{\psi}(\vec{k})|^{2}$ over a large
number $M$ of different filter functions $f_{i}$. Since the mean value
of $\phi_{i}(\vec{k})$ is zero, this averaging results in an
approximation to $|\psi(\vec{k})|^{2}$ as an incoherent sum of the two
terms,
\begin{equation}
  \langle|\tilde{\psi}(\vec{k})|^{2}\rangle_{M}\approx\mathcal{N'}\bigg[|\psi(\vec{k})|^{2} + \frac{1}{NP}\langle|(\psi(\vec{k})\ast\phi_{i}(\vec{k}))|^{2}\rangle_{M}\bigg],
\end{equation}
where $\langle\cdot\rangle_{M}$ is an average over $M$ filter
functions.

\section{Compressive Sensing}

Compressive sensing (CS) is a measurement technique that uses
optimization to obtain a $N$-dimensional signal $\bm{X}$ from $M<<N$
linear projections (linear
measurements) \cite{baraniuk:2007,candes:2006}. CS exploits
prior-knowledge about the signal's compressibility to require fewer
measurements than the Nyquist limit. The measurement process is
\begin{equation}
\bm{Y} = \bm{FX} + \bm{\Gamma},
\end{equation}
where $\bm{Y}$ is an $M$-dimensional vector of measurements, $\bm{F}$
is an $M\times N$ sensing matrix, and $\bm{\Gamma}$ is an
$M$-dimensional noise vector. Each measured value $\bm{Y}_i$ is
therefore the inner-product of $\bm{X}$ with sensing vector
$\bm{F}_i$, where $i$ is an index over rows of $\bm{F}$.

Because $M<<N$, $\bm{Y}$ does not uniquely specify $\bm{X}$. CS
proposes that the correct $\bm{X}$ is the one which is most
compressible by a method expected to compress it. Most commonly, one
must know a basis or transformation in which $\bm{X}$ is expected to
be sparse (have few nonzero coefficients). For images, typical sparse
representations include discrete cosines, various wavelets, and the
discrete gradient \cite{romberg:2008}.

This correct $\bm{X}$ is found by minimizing the objective function
\begin{equation}
  \min_{\bm{X}} \frac{\mu}{2}||\bm{Y}-\bm{FX}||_2^2 + g(\bm{X}),
  \label{eq:obj}
\end{equation}
where for example $||\bm{Q}||_2^2$ is the $\ell_2$ (Euclidean) norm of
$\bm{Q}$ and $\mu$ is a scalar constant. The first penalty is a
least-squares penalty; it ensures the recovered $\bm{X}$ is consistent
with the measurements. The second penalty $g(\bm{X})$ is a term which
gets smaller the more compressible $\bm{X}$ is. Typical $g(\bm{X})$
include the $\ell_1$ norm of $\bm{\Phi X}$
\begin{equation}
  g(\bm{X}) = ||\bm{\Phi} \bm{X}||_1= \sum_{i=1}^{M}|\bm{\Phi} \bm{X}|,
\end{equation}
where $\bm{\Phi}$ is a transform to a sparse basis (wavelets,
cosines), and $\bm{X}$'s total variation
\begin{equation}
   g(\bm{X}) = TV(\bm{X}) = \sum_{\text{adj. }i,j}|\bm{X}_i-\bm{X}_j|,
\end{equation}
where $i$ and $j$ run over pairs of adjacent pixels in $\bm{X}$. This
is the $\ell_1$ norm of $\bm{X}$'s discrete gradient
\cite{sup:wang:2008}. The $\ell_1$ norm is a useful measure of sparsity
because it makes Eq. \ref{eq:obj} convex and therefore easy to solve.

To minimize the required number of measurements $M$, the sensing
vectors $\bm{F}_i$ should be mutually unbiased with the sparse
transform; this gives the counter-intuitive result that random sensing
vectors are extremely effective in almost all cases. For a $K$-sparse
signal ($K$ nonzero entries in the sparse representation), CS can give
an exact reconstruction with only $M \propto K\log(N/K)$ measurements
\cite{sup:candes:2006:2}. In practice, $M$ can be as small as a few
percent of $N$.

\subsection{\label{sec:Pix}Single-Pixel Camera}
The most illustrative example of compressive sensing is the Rice
single-pixel camera \cite{baraniuk:2008}. The camera is composed of a
single pixel detector, a digital micro-mirror device (DMD), and an
imaging lens. A DMD is a mirror array composed of thousands of
mirrors. Each mirror acts as a reflective pixel with values on and
off, reflecting into and away from the single-pixel detector
respectively. The lens images an object onto the DMD array while the
DMD displays a random pattern corresponding to a row in the sensing
matrix $\bm{F}$. The single-pixel detector records the intensity of
light as a projection of the random pattern with the object for all
$M$ patterns resulting in a $\bm{Y}$ vector of length $M$. The sensing
matrix $\bm{F}$ and the measurement vector $\bm{Y}$ are fed into an
algorithm to minimize equation (\ref{eq:obj}).



\end{document}